\documentclass[twocolumn,superscriptaddress,showpacs,nofootinbib,preprintnumbers,secnumarabic,amssymb, nobibnotes, aps, prd]{revtex4-2}
\usepackage[utf8]{inputenc}
\usepackage{graphicx}
\usepackage{latexsym,amsmath,amssymb,amsthm,lmodern,float,url}
\usepackage{natbib}
\usepackage{color}
\usepackage{microtype}
\usepackage{import}
\usepackage{bbold}
\usepackage[plain]{fancyref}
\usepackage{varioref}
\usepackage{slashed}
\usepackage{multirow}
\usepackage{tikz}
\usepackage{scrextend}
\usepackage{braket}
\usetikzlibrary{shapes}
\usetikzlibrary{positioning}
\usepackage[normalem]{ulem}
\usepackage{caption}
\usepackage{subcaption}
\usepackage{units}
\usepackage{qcircuit}
\usepackage{enumitem}

\newcommand{\prep}{\mathcal{A}}
\newcommand{\grov}{\mathcal{Q}}
\newcommand{\qop}{\mathbb{V}}

\newcommand{\Mod}{\text{ mod }}

  \usepackage{float}
  \usepackage{afterpage}
  \usepackage{array}
  \usepackage{booktabs}
  \usepackage{multirow}
  \usepackage{tabularx}
  
  \newcolumntype{P}[1]{>{\centering\arraybackslash}p{#1}} 
  \newcolumntype{M}[1]{>{\centering\arraybackslash}m{#1}} 
  \newcolumntype{B}[1]{>{\centering\arraybackslash}b{#1}} 
  
  \usepackage{dcolumn}
  \newcolumntype{.}{D{.}{.}{-1}} 

\usepackage[colorlinks=true,backref=false, linktocpage=true,
citecolor=blue,urlcolor=blue,linkcolor=blue,pdfpagemode=UseOutlines]{hyperref}
\hypersetup{%
  bookmarksnumbered=true,
  pdftitle = {},
  pdfsubject = {},
  pdfauthor = {},
  pdfkeywords = {}
}

\usepackage{orcidlink}

\let\Re\undefined
\let\Im\undefined
\DeclareMathOperator{\Tr}{Tr}
\DeclareMathOperator{\Re}{Re}
\DeclareMathOperator{\Im}{Im}
\DeclareMathOperator{\tr}{Tr}

\newtheorem{theorem}{Theorem}[section]

\let\originalleft\left
\let\originalright\right
\renewcommand{\left}{\mathopen{}\mathclose\bgroup\originalleft}
\renewcommand{\right}{\aftergroup\egroup\originalright}

\newcommand{\ceil}[1]{\lceil #1 \rceil}

\begin{document}
\preprint{FERMILAB-PUB-24-0264-T}
\title{Fermion determinants on a quantum computer}

\author{George T.\ Fleming \orcidlink{0000-0002-4987-7167}}
\email{gfleming@fnal.gov}
\author{Prasanth Shyamsundar \orcidlink{0000-0002-2748-9091}}
\email{prasanth@fnal.gov}
\author{Judah Unmuth-Yockey \orcidlink{0000-0001-9962-7134}}
\email{jfunmuthyockey@gmail.com}
\affiliation{Fermi National Accelerator Laboratory, Batavia,  Illinois, 60510, USA}

\date{\today}

\begin{abstract}
We present a quantum algorithm to compute the logarithm of the determinant of the fermion matrix, assuming access to a classical lattice gauge field configuration.  The algorithm uses the quantum eigenvalue transform, and quantum mean estimation, giving a query complexity that scales like $O(V\log(V))$ in the matrix dimension $V$.
\end{abstract}

\maketitle

\section{Introduction}

Lattice quantum chromodynamics (QCD) is a cornerstone in modern high-energy physics theory.  Under a Wick rotation, the Minkowski signature of spacetime is transformed into Euclidean space allowing for the path integral of QCD to be interpreted as a classical partition function in four Euclidean spatial dimensions (see Refs.~\cite{montvay_munster_1994,gattringer,kogut:1979} for review). With the discretization of this space into a lattice, this allows for the Monte Carlo method to be used to sample that partition function.  However, with the inclusion of fermionic fields into the lattice action in the form of Grassmann variables, this sampling cannot be done directly, and instead the fermions must be integrated out beforehand
which results in an effective action of the gauge fields which depends on the logarithm of the determinant of the fermion matrix, $M$.  Therefore, to sample the QCD partition function, the determinant of the fermion matrix must be calculated during each Monte Carlo step.  This calculation is expensive, scaling like $O(V^{3})$, where $V$ is the dimension of the fermion matrix, and hence, finding improvements to this scaling would assist in lattice QCD calculations.

The standard method to avoid this expensive calculation is to introduce new bosonic fields called pseudofermions \cite{WEINGARTEN1981333, Gottlieb:1987mq} which replaces the problem of computing the log of the determinant with the problem of solving a sparse linear system which converges faster than $O(V^{3})$. The pseudofermion method places limitations on the types of physical systems that can be solved.  In particular, finite density calculations suffer from a severe sign problem which make it difficult to perform calculations relevant for heavy-ion collision experiments and the equation of state for dense nuclear matter in the heart of neutron stars. Having access to an efficient algorithm for computing log of the determinant would enable calculations at finite density free of sign problems \cite{Joo:2001bz,PhysRevD.105.L051506,Nagata_2022}.

Aside from the problem of efficient sampling of the lattice QCD partition function, computation of physical observables involving fermions on the sampled configurations often involve computing the trace of the inverse of the fermion matrix $M^{-1}$ which is also called the fermion propagator.  There are many areas of particle physics research that can benefit from more accurate and more frequent computation of these traces including neutral current neutrino-nucleus scattering in the DUNE and SBN experimental programs \cite{Park:2024vjp}, quark-disconnected contributions to the anomalous magenetic moment of the muon related to the Fermilab $g-2$ experiment \cite{Kuberski:2023qgx}, couplings of Higgs bosons to nuclei relevant for direct-detection of dark matter scattering off normal matter through Higgs exchange \cite{Ellis:2018dmb, Varnhorst:2020dba}, and for computation of properties of composite Higgs bosons in models which may be relevant to the Large Hadron Collider (LHC) \cite{LatticeStrongDynamics:2023bqp}, just to name a few.

Quantum computing offers many potential speed-ups to classical algorithms.  One of the most promising is through a generic algorithm called the quantum eigenvalue (or singular value) transform (QET)~\cite{PRXQuantum.2.040203,low2024quantum,Gily_n_2019}.  The strategy of this algorithm is to construct polynomial transforms of matrices using alternating $X$- and $Z$-like rotations on a two-dimensional qubit subspace.  The depth of these quantum circuits for these transforms scale as the degree of the polynomial being implemented, providing exponential improvements over classical algorithms in many cases, and often providing optimal constructions~\cite{low2024quantum,Gily_n_2019}.
Another area where quantum computers offer an advantage is in the calculation of normalized sums (or means).  Quantum mean estimation (QME) is known to provide a quadratic speed-up over its classical counterpart~\cite{kothari:2022,shyamsundar2021nonboolean,Ham21,montanaro:2015}.

Using the aforementioned technology, we propose to accelerate the calculation of fermion determinants. The starting point is the following well-known identity for a positive-definite matrix $W$: $\det[W] = \exp(\Tr[\log W])$. Our approach is to a) use the QET to compute an approximation to $\log W$ and subsequently b) use the QME algorithm to compute the trace, in order to estimate $\det[W]$. Using this strategy to evaluate the determinant of the fermion matrix $M$ involves the following ``classical'' steps as groundwork: 1) Devise a positive-definite matrix $W$ whose determinant has a one-to-one correspondence to the determinant of $M$ (which is not necessarily positive-definite). 2) Construct an efficient quantum circuit to block-encode \cite{Low_2019,PhysRevLett.118.010501,camps2023explicit} the matrix $W$; this is necessary to perform the QET on $W$. 3) Customize the implementation of QET so that the corresponding transformation approximates the $\log$ function.
If the computational cost of the resulting quantum algorithm is found to be less than $O(V^{3})$ there is a potential speed up of calculating fermion determinants using a quantum computer.  Indeed we show that in terms of the polynomial degree $d$, the probability of success $1-\delta$, the error on the trace estimation $\epsilon$, and the dimension of the matrix $V$, the algorithm's query complexity scales like $O(d \log(1/\delta) V \log(V) / \epsilon)$.

The article is organized as follows: In Sec.~\ref{sec:block-encoding} we devise block-encodings for staggered fermions coupled to an SU(3) gauge field, starting from the simple case of a free scalar field Laplacian.  In Sec.~\ref{sec:matrix-trace} we show how to compute the trace of a unitary matrix, as well as the specialization to computing the trace of a block-encoded matrix.  Then in Sec.~\ref{sec:matrix-det} we combine the previous ingredients in the context of the QET to demonstrate how to compute $\log\det(M)$.  Finally in Sec.~\ref{sec:conclusion} we give some concluding remarks as well as further optimizations and future work to be done.

\section{Block-encoding}
\label{sec:block-encoding}
In order to use the quantum computer to compute the determinant
of the fermion matrix, we must
first load the fermion matrix onto the quantum computer.  The fermion matrix itself is not unitary, and so
to do this we encode it inside a larger unitary matrix through the method
of block-encoding.
To block-encode a matrix, one embeds the nonunitary matrix in the top-left block of a larger matrix, and from that block determines what the remaining entries must be such that the larger matrix is unitary, as is required in quantum computing.  For example, a simple block-encoding of a Hermitian matrix $A$ into a larger matrix $B$ is given by,
  \begin{align}
    B =
    \begin{pmatrix}
      A & \sqrt{1 - A^{2}} \\
      -\sqrt{1 - A^{2}} & A
    \end{pmatrix}.
  \end{align}
  For a generic block-encoding which uses $\ell+1$ qubits in which to embed the block, this can be expressed as
  \begin{align}
    \label{eq:block-def}
    A = (\bra{0^{\ell+1}} \otimes \mathbb{1}) B (\ket{0^{\ell+1}} \otimes \mathbb{1}).
  \end{align}

Rather than immediately providing the block-encoding of the fermion matrix, it is
advantageous to build up the block-encoding process for simpler cases which at each step provide necessary technology needed
in order to block encode the full fermion matrix.  We begin with a free scalar field Laplacian, then free staggered
fermions, staggered fermions with abelian gauge fields, and then finally SU(3) gauge fields with staggered fermions.
Since these matrices are sparse, we use the methods of Ref.~\cite{camps2023explicit} to block-encode them.

\subsection{Scalar Laplacian}
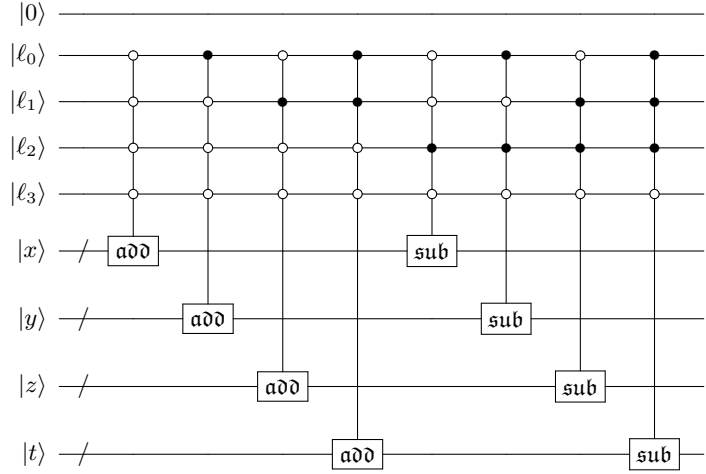
\begin{figure*}
    \centering
    \mbox{
    \Qcircuit @C=1em @R=1.5em {
    \lstick{\ket{0}} & \qw & \qw & \qw & \qw & \qw & \qw & \qw & \qw & \qw & \qw \\
    \lstick{\ket{\ell_{0}}} &  \qw & \ctrlo{1}  & \ctrl{1} & \ctrlo{1} & \ctrl{1}
    & \ctrlo{1}  & \ctrl{1} & \ctrlo{1} & \ctrl{1} & \qw \\
    \lstick{\ket{\ell_{1}}} &  \qw & \ctrlo{1} & \ctrlo{1}  & \ctrl{1} & \ctrl{1}
    & \ctrlo{1} & \ctrlo{1}  & \ctrl{1} & \ctrl{1} & \qw \\
    \lstick{\ket{\ell_{2}}} &  \qw & \ctrlo{1} & \ctrlo{1} & \ctrlo{1}  & \ctrlo{1}
    & \ctrl{1} & \ctrl{1}  & \ctrl{1} & \ctrl{1} & \qw \\
    \lstick{\ket{\ell_{3}}} &  \qw & \ctrlo{1} & \ctrlo{2} & \ctrlo{3} & \ctrlo{4}
    & \ctrlo{1} & \ctrlo{2}  & \ctrlo{3} & \ctrlo{4} & \qw \\
    \lstick{\ket{x}} & {/} \qw & \gate{\mathfrak{add}} & \qw & \qw & \qw & \gate{\mathfrak{sub}} 
    & \qw & \qw & \qw & \qw \\
    \lstick{\ket{y}} & {/} \qw & \qw & \gate{\mathfrak{add}} & \qw & \qw & \qw
    & \gate{\mathfrak{sub}} & \qw & \qw & \qw \\
    \lstick{\ket{z}} & {/} \qw & \qw & \qw & \gate{\mathfrak{add}} & \qw & \qw
    & \qw & \gate{\mathfrak{sub}} & \qw & \qw \\
    \lstick{\ket{t}} & {/} \qw & \qw & \qw & \qw & \gate{\mathfrak{add}} & \qw
    & \qw & \qw & \gate{\mathfrak{sub}} & \qw 
    }}
    \caption{$O^{NN}_{c}$ operator for nonzero matrix elements between nearest-neighbors on the lattice.}
    \label{fig:oc}
\end{figure*}

\begin{figure}
    \centering
    \mbox{
    \Qcircuit @C=1em @R=1.5em {
    \lstick{\ket{0}} & \qw & \gate{R_{y}(\theta_{1})} & \gate{R_{y}(\theta_{0})} & \qw \\
    \lstick{\ket{\ell_{0}}} &  \qw & \qw & \qw
     & \qw \\
    \lstick{\ket{\ell_{1}}} &  \qw & \qw & \qw
    & \qw \\
    \lstick{\ket{\ell_{2}}} &  \qw & \qw & \qw
    & \qw \\
    \lstick{\ket{\ell_{3}}} &  \qw & \ctrlo{-4} & \ctrl{-4}
    & \qw \\
    \lstick{\ket{\vec{n}}} & {/} \qw & \qw & \qw & \qw 
    }}
    \caption{$O^{s}_{A}$ operator for the scalar Laplacian.  Filling in the nine nonzero entries.}
    \label{fig:oa}
\end{figure}
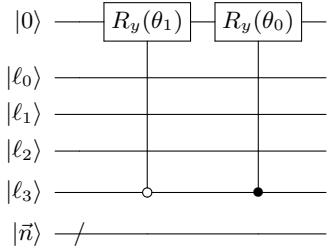

\begin{figure}
    \centering
    \mbox{
    \Qcircuit @C=1em @R=1.5em {
    \lstick{\ket{0}} & \qw & \qw & \multigate{1}{O^{s}_{A}} & \qw & \qw & \qw \\
    \lstick{\ket{\ell}} & {/}  \qw & \gate{D_{s}} & \ghost{O^{s}_{A}}  & \multigate{1}{O_{c}} & \gate{D_{s}} & \qw \\
    \lstick{\ket{\vec{n}}} & {/}  \qw & \qw & \qw & \ghost{O_{c}}  & \qw & \qw
    }}
    \caption{The block-encoding circuit for the free scalar Laplacian.}
    \label{fig:block}
\end{figure}
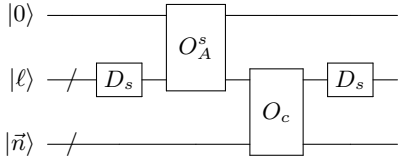

We consider a normalized lattice Laplacian,
\begin{align}
  \nonumber
    \Delta &= \frac{1}{s}(-\nabla^{2} + m_{0}^{2}) \\
    &= \frac{1}{s}
    \begin{pmatrix}
        q+m_{0}^{2} & \cdots & -1 & \cdots  \\
        0 & \ddots &  0 & \cdots \\
        \vdots & 0 & \ddots & \vdots  \\
        \cdots & -1 & \cdots & q+m_{0}^{2} 
    \end{pmatrix},
\end{align}
where $q = 2D$ is the coordination number of the hypercubic lattice in $D$ dimensions, and $m_{0}$ is the bare mass.  Each row/column, which corresponds to a lattice site, has two unique values: $\{-1, q+m_{0}^{2}\}$, and $2D+1$ nonzero entries total.

Let us fix $D=4$.  We can write the list of nonzero elements in 
a given column corresponding to the lattice site $\vec{n}$
as $N = \{-1, -1, -1, -1, -1, -1, -1, -1, 8+m_{0}^{2}\} / s$ where the first eight entries lie in rows which correspond to $\{\vec{n}\pm\hat{x}, \vec{n}\pm\hat{y}, \vec{n}\pm\hat{z}, \vec{n}\pm\hat{t}\}$. Following Ref.~\cite{camps2023explicit}, for each column $j$, we need to map each non-zero entry to one or more values of an index $\ell$. We will use $\ell=0$-$7$ to encode the the terms corresponding to the eight nearest neighbor elements and $8\leq \ell<16$ to encode the diagonal element. Let $c(j, \ell)$ represent the row-index of the element on the $j$-th column associated with index $\ell$,
\begin{align}
\label{eq:scalar_cjl}
    c(\vec{n},\ell) = 
    \begin{cases}
        \ell = 0 : & \vec{n} + \hat{x} \\
        \ell = 1 : & \vec{n} + \hat{y} \\
        \ell = 2 : & \vec{n} + \hat{z} \\
        \ell = 3 : & \vec{n} + \hat{t} \\
        \ell = 4 : & \vec{n} - \hat{x} \\
        \ell = 5 : & \vec{n} - \hat{y} \\
        \ell = 6 : & \vec{n} - \hat{z} \\
        \ell = 7 : & \vec{n} - \hat{t} \\
        8 \leq \ell < 16 : & \vec{n} \\
    \end{cases}.
\end{align}
An operator $O^{NN}_{c}$ which implements this transformation is
\begin{align}
    O^{NN}_{c} \ket{0000} \ket{\vec{n}} &= \ket{0000} \ket{(x+1)\Mod L}\ket{y}\ket{z}\ket{t} \\
    O^{NN}_{c} \ket{0001} \ket{\vec{n}} &= \ket{0001} \ket{x}\ket{(y+1) \Mod L}\ket{z}\ket{t} \\
    \vdots \nonumber \\
  O^{NN}_{c} \ket{0111} \ket{\vec{n}} &= \ket{0111} \ket{x}\ket{y}\ket{z}\ket{(t-1) \Mod L} \\
  O^{NN}_{c} \ket{1abc} \ket{\vec{n}} &= \ket{1abc} \ket{x}\ket{y}\ket{z}\ket{t},\quad \forall a,b,c\in\{0,1\}
\end{align}
The operator $O^{NN}_{c}$ only depends on $\ell$, and so we can perform controlled integer arithmetic using the $\ell$ register to implement $O^{NN}_{c}$.  The modular addition and subtraction operators needed above are given by
\begin{align}
    \mathfrak{add} = 
    \begin{pmatrix}
        0 & 0 & \ldots & 0 & 1 \\
        1 & 0 & \ddots & \ddots & 0 \\
        0 & 1 & \ddots & \ddots & 0 \\
        \vdots & \ddots & \ddots & \ddots & \vdots \\
        0 & 0 & \ldots & 1 & 0
    \end{pmatrix}
\end{align}
and $\mathfrak{sub} = \mathfrak{add}^{T}$.  Figure~\ref{fig:oc} shows the circuit for the $O^{NN}_{c}$ operator which connects nearest neighbors.

The other ingredient is an operator $O^{s}_{A}$ which gives
\begin{align}
    O^{s}_{A} \ket{0} \ket{\ell} \ket{\vec{n}} &= 
  \left( A_{c(\vec{n},\ell) \vec{n}} \ket{0} \right. \\ \nonumber
  &+ \left. \sqrt{1 - |A_{c(\vec{n},\ell) \vec{n}}|^{2}} \ket{1} \right) \ket{\ell} \ket{\vec{n}},
\end{align}
where $A_{c(\vec{n},\ell) \vec{n}}$ is the matrix element at ${c(\vec{n},\ell), \vec{n}}$.  The value that $O_{A}$ assigns is determined by $\ell$ only.  So we only need controlled rotations---controlled on the $\ell$ register acting on the $\ket{0}$ register---to implement $O^{s}_{A}$. Figure~\ref{fig:oa} gives the circuit for this operator.  In this case when the matrix elements are themselves real, we can use a $y$-rotation gate.

We also use $D_{s}$, the ``diffusion operator'', which creates an equal superposition on the qubits in the $\ket{\ell}$-register using Hadamards.  The complete circuit for the block encoding of the free scalar Laplacian can be seen in Fig.~\ref{fig:block}.  The following block-encodings have a similar complete structure.

To compute the angles for the rotations in $O_{A}^{s}$ we look at expectation values.  For the diagonal,
\begin{align}
  \label{eq:scalar-diagonal}
  \nonumber
  &\bra{\vec{n}} \bra{0000}  \bra{0} D_{s} O_{c}^{s} O_{A}^{s} D_{s} \ket{0} \ket{0000} \ket{\vec{n}} \\ \nonumber
  &= \frac{1}{16} (8\cos(\theta_{0}/2)) \equiv \frac{1}{s}(8 + m_{0}^{2}) \\
  &\implies \theta_{0} = 2 \arccos\left(\frac{2}{s}(8 + m_{0}^{2}) \right)
\end{align}
For any one of the off-diagonal elements we consider
\begin{align}
  \label{eq:scalar-offd}
  \nonumber
  &\bra{\vec{n}+\hat{\mu}} \bra{0000} \bra{0} D_{s} O_{c}^{s} O_{A}^{s} D_{s} \ket{0} \ket{0000} \ket{\vec{n}} \\ \nonumber
  &= \frac{1}{16} \cos(\theta_{1}/2) \equiv -\frac{1}{s} \\
  &\implies \theta_{1} = 2 \arccos(-16/s).
\end{align}
We see from Eqs.~\eqref{eq:scalar-diagonal} and~\eqref{eq:scalar-offd} how to choose the normalization so that the argument of arccosine is valid.  In this case $s$ must be chosen such that
\begin{align}
  \label{eq:scalar-s-norm}
  |s| \geq 2\,(8+m_{0}^{2}), \quad \text{and} \quad |s| \geq 16.
\end{align}
Similar arguments can be used for the following cases.

\subsection{Free staggered fermions}
\label{sec:free-fermions}

We now move on to the block-encoding of free staggered fermions.
The action is given by,
\begin{align}
    S = \sum_{n} \bar{\chi}_{n} \left( K \sum_{\mu=1}^{4} 
    \eta_{\mu}(n) \Delta_{\mu} + m_{0}   \right) \chi_{n}
\end{align}
with
\begin{align}
    \Delta_{\mu} \chi_{n} = \frac{\chi_{n+\hat{\mu}} - \chi_{n-\hat{\mu}}}{2}
\end{align}
as the symmetric finite difference, $\eta_{\mu}(n) = (-1)^{\sum_{\nu < \mu} n_{\nu}}$ with $\eta_{1}(n) = 1$ is the staggered phase, $K$
and $m_{0}$ are couplings, and $\bar{\chi}$ and $\chi$ are Grassmann fields.  The $K$ coupling is typically set to one, but we leave it general here.

Write the free staggered fermion matrix as
\begin{align}
  \label{eq:free-stag-matrix}
  M_{m n} = \frac{K}{2} \sum_{\mu} \eta_{\mu}(m) (\delta_{n, m+\hat{\mu}} - \delta_{n, m-\hat{\mu}}) + m_{0} \delta_{n, m}
\end{align}
so that
\begin{align}
  \label{eq:short-free-act}
  S = \sum_{m,n} \bar{\chi}_{m} M_{m n} \chi_{n}.
\end{align}

At this point it is useful to reflect on some of the important properties of $M$, and how we wish to compute the determinant of $M$.  The goal is to use the relation $\Tr[\log] = \log\det$ to compute the trace of the log of a matrix and avoid ever computing the actual determinant.  However, the logarithm is only defined for positive arguments, and $M$ generally can have complex eigenvalues making the direct logarithm of $M$ undefined.  To resolve this issue, we use the nice property of $M$ that its eigenvalues always appear in complex conjugate pairs (see appendix~\ref{sec:app:ferm-eig-relations}) .  This results in the determinant of $M$ as being purely real and positive.  This in turn gives rise to the following relation between $M$ and $W$:
\begin{align}
  \label{eq:w-def}
  W_{a c} \equiv \sum_{b} M^{\dagger}_{a b} M_{b c} = \sum_{b} M^{*}_{b a} M_{b c},
\end{align}
giving
\begin{align}
  \label{eq:w-mm}
  2\log\det(M) = \log\det(W).
\end{align}
$W$ \emph{is} positive definite, meaning that the logarithm of that matrix is well-defined.  We then block-encode $W$, instead of $M$, knowing that when we compute the trace of the logarithm the outcome only differs from the desired answer by a factor of two.

Using Eq.~\eqref{eq:w-def}, we can work out $W$, giving,
\begin{align}
  \nonumber
  W_{a c} &= \frac{K^{2}}{4}  \sum_{\mu, \nu} \left[
            \eta_{\mu}(a+\hat{\mu}) \eta_{\nu}(c+\hat{\nu}) \delta_{a+\hat{\mu}-\hat{\nu}, c} \right. \\ \nonumber
          &- \eta_{\mu}(a+\hat{\mu}) \eta_{\nu}(c-\hat{\nu}) \delta_{a+\hat{\mu}+\hat{\nu}, c} \\ \nonumber
          &- \eta_{\mu}(a-\hat{\mu}) \eta_{\nu}(c+\hat{\nu}) \delta_{a-\hat{\mu}-\hat{\nu}, c} \\ \nonumber
          & \left. + \eta_{\mu}(a-\hat{\mu}) \eta_{\nu}(c-\hat{\nu}) \delta_{a-\hat{\mu}+\hat{\nu}, c} \right] \\ \nonumber
          &+ m_{0} \frac{K}{2} \sum_{\mu} \eta_{\mu}(c) (\delta_{c, a+\hat{\mu}} - \delta_{c, a-\hat{\mu}}) \\ \nonumber
          &+ m_{0} \frac{K}{2} \sum_{\nu} \eta_{\nu}(a) (\delta_{a, c+\hat{\nu}} - \delta_{a, c-\hat{\nu}}) \\
          &+ m_{0}^{2} \delta_{a, c}.
\end{align}
The staggered phase has the property that shifting $\eta_{\mu}(x)$ by any $\nu \geq \mu$ doesn't change it.  Therefore,
\begin{align}
  \label{eq:staggered_property}
  -\eta_{\mu}(a+\hat{\mu}) + \eta_{\mu}(a) = 0
\end{align}
and the terms connecting nearest neighbors cancel.  Thus,
\begin{align}
  \label{eq:first-free-W}
  \nonumber
  W_{a c} &= \frac{K^{2}}{4}  \sum_{\mu, \nu} \eta_{\mu}(a) \eta_{\nu}(c) \left[
             \delta_{a+\hat{\mu}-\hat{\nu}, c} \right. \\ \nonumber
          &-  \delta_{a+\hat{\mu}+\hat{\nu}, c}
          -  \delta_{a-\hat{\mu}-\hat{\nu}, c} \\
          & \left. +  \delta_{a-\hat{\mu}+\hat{\nu}, c} \right]
          + m_{0}^{2} \delta_{a, c}.
\end{align}
We see this matrix only connects those lattice sites that are two ``hops'' away, or next-to-nearest neighbors (NtNNs).

We can enumerate those neighbors to know how many nonzero elements there are in each column of $W$.  Choose an origin lattice site.  There are four-choose-two combinations of $\mu$ and $\nu$ defining six planes through the origin, each containing four sites that are NtNNs.  This gives $6 \times 4 = 24$ NtNNs.  There are then those points in each of the eight directions away from the origin that are two hops away, adding eight more sites, giving $24 + 8 = 32$ NtNNs total (there is also a set of two hops which return to the origin.  There are eight of these).  We can re-express Eq.~\eqref{eq:first-free-W} specifically identifying these NtNN sites,
\begin{align}
  \label{eq:free-W}
  \nonumber
  W_{a c} &= \frac{K^{2}}{4}  \sum_{\mu < \nu} (\eta_{\mu}(a) \eta_{\nu}(c) + \eta_{\nu}(a) \eta_{\mu}(c)) [
            \delta_{a+\hat{\mu}-\hat{\nu}, c} \\ \nonumber
          &+ \delta_{a-\hat{\mu}+\hat{\nu}, c}
            -  \delta_{a+\hat{\mu}+\hat{\nu}, c} \\ \nonumber
          &-  \delta_{a-\hat{\mu}-\hat{\nu}, c} ]  + (m_{0}^{2} + 2K^{2}) \delta_{a, c} \\
          &- \frac{K^{2}}{4} \sum_{\mu} (\delta_{a+2\hat{\mu},c} + \delta_{a-2\hat{\mu}, c}).
\end{align}
In addition, another property of the staggered phases is that when $\mu < \nu$, then $\eta_{\nu}(x \pm \mu) + \eta_{\nu}(x) = 0$.  This property completely eliminates the first term in Eq.~\eqref{eq:free-W}, leaving only the last two terms:
\begin{align}
  \label{eq:final-free-W}
  W_{a c}  &= (m_{0}^{2} + 2K^{2}) \delta_{a, c} \\ \nonumber
          &- \frac{K^{2}}{4} \sum_{\mu} (\delta_{a+2\hat{\mu},c} + \delta_{a-2\hat{\mu}, c}).
\end{align}

The $O^{f}_{c}$ operator in this case is almost identical to the free scalar case with $O_{c}^{NN}$ with the replacement $\mathfrak{add} \rightarrow \mathfrak{add}^{2}$, and $\mathfrak{sub} \rightarrow \mathfrak{sub}^{2}$.  This programs two hops in each of the eight directions instead of one.
For $O_{A}^{f}$ we can also use a similar structure to $O_{A}^{s}$.  Since there are only nine nonzero entries in each row/column, and they are real numbers, we can encode them using $y$-rotations like Fig.~\ref{fig:oa}.  If we block-encode $W / s$ for some normalization $s$, we find the angles are given by
\begin{align}
  \label{eq:free-angles-diag}
  \theta_{0} = 2 \arccos\left( \frac{2}{s} (m_{0}^{2} + 2K^{2}) \right)
\end{align}
for the diagonal, and
\begin{align}
  \label{eq:free-angles-offdiag}
  \theta_{1} = 2 \arccos\left(-\frac{4 K^{2}}{s} \right)
\end{align}
for the off-diagonal terms.

\subsection{U(1) and staggered fermions}
\label{sec:u1-ferms}
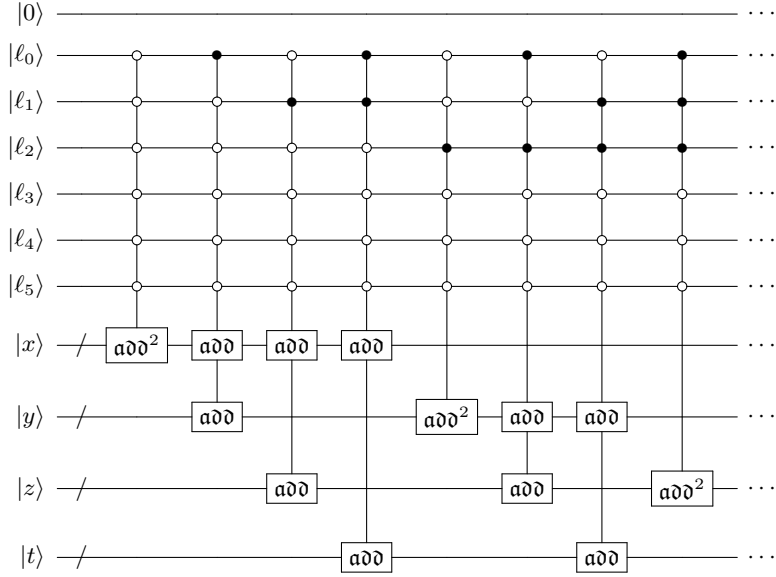
\begin{figure*}
    \centering
    \mbox{
    \Qcircuit @C=1em @R=1.5em {
    \lstick{\ket{0}} & \qw & \qw & \qw & \qw & \qw & \qw & \qw & \qw & \qw & \qw & \cdots \\
    \lstick{\ket{\ell_{0}}} &  \qw & \ctrlo{1}  & \ctrl{1} & \ctrlo{1} & \ctrl{1}
    & \ctrlo{1}  & \ctrl{1} & \ctrlo{1} & \ctrl{1} & \qw & \cdots \\
    \lstick{\ket{\ell_{1}}} &  \qw & \ctrlo{1} & \ctrlo{1}  & \ctrl{1} & \ctrl{1}
    & \ctrlo{1} & \ctrlo{1}  & \ctrl{1} & \ctrl{1} & \qw & \cdots \\
    \lstick{\ket{\ell_{2}}} &  \qw & \ctrlo{1} & \ctrlo{1} & \ctrlo{1}  & \ctrlo{1}
    & \ctrl{1} & \ctrl{1}  & \ctrl{1} & \ctrl{1} & \qw & \cdots \\
    \lstick{\ket{\ell_{3}}} &  \qw & \ctrlo{1} & \ctrlo{1} & \ctrlo{1} & \ctrlo{1}
    & \ctrlo{1} & \ctrlo{1}  & \ctrlo{1} & \ctrlo{1} & \qw & \cdots \\
    \lstick{\ket{\ell_{4}}} &  \qw & \ctrlo{1} & \ctrlo{1} & \ctrlo{1} & \ctrlo{1}
    & \ctrlo{1} & \ctrlo{1}  & \ctrlo{1} & \ctrlo{1} & \qw & \cdots \\
    \lstick{\ket{\ell_{5}}} &  \qw & \ctrlo{1} & \ctrlo{1} & \ctrlo{1} & \ctrlo{1}
    & \ctrlo{2} & \ctrlo{2}  & \ctrlo{2} & \ctrlo{3} & \qw & \cdots \\
    \lstick{\ket{x}} & {/} \qw & \gate{\mathfrak{add}^{2}} & \sgate{\mathfrak{add}}{1} & \sgate{\mathfrak{add}}{2} & \sgate{\mathfrak{add}}{3} & \qw
    & \qw & \qw & \qw & \qw & \cdots \\
    \lstick{\ket{y}} & {/} \qw & \qw & \gate{\mathfrak{add}} & \qw & \qw & \gate{\mathfrak{add}^{2}}
    & \sgate{\mathfrak{add}}{1} & \sgate{\mathfrak{add}}{2} & \qw & \qw & \cdots \\
    \lstick{\ket{z}} & {/} \qw & \qw & \qw & \gate{\mathfrak{add}} & \qw & \qw
    & \gate{\mathfrak{add}} & \qw & \gate{\mathfrak{add}^{2}} & \qw & \cdots \\
    \lstick{\ket{t}} & {/} \qw & \qw & \qw & \qw & \gate{\mathfrak{add}} & \qw
    & \qw & \gate{\mathfrak{add}} & \qw & \qw  & \cdots
    }}
    \caption{The start of the $O^{NtNN}_{c}$ operator for the nonzero matrix elements between NtNNs on the lattice.}
    \label{fig:ocntnn}
  \end{figure*}
\begin{figure*}
    \centering
    \mbox{
    \Qcircuit @C=1em @R=1.5em {
    \lstick{\ket{0}} & \qw & \gate{U_{3}(\vec{\theta}_{\vec{n},0})} & \gate{U_{3}(\vec{\theta}_{\vec{n},1})} & \gate{U_{3}(\vec{\theta}_{\vec{n},2})} & \gate{U_{3}(\vec{\theta}_{\vec{n},3})} & \gate{U_{3}(\vec{\theta}_{\vec{n},4})} & \gate{U_{3}(\vec{\theta}_{\vec{n},5})} & \gate{U_{3}(\vec{\theta}_{\vec{n},6})} & \gate{U_{3}(\vec{\theta}_{\vec{n},7})} & \qw & \cdots \\
    \lstick{\ket{\ell_{0}}} &  \qw & \ctrlo{-1}  & \ctrl{-1} & \ctrlo{-1} & \ctrl{-1}
    & \ctrlo{-1}  & \ctrl{-1} & \ctrlo{-1} & \ctrl{-1} & \qw & \cdots \\
    \lstick{\ket{\ell_{1}}} &  \qw & \ctrlo{-1} & \ctrlo{-1}  & \ctrl{-1} & \ctrl{-1}
    & \ctrlo{-1} & \ctrlo{-1}  & \ctrl{-1} & \ctrl{-1} & \qw & \cdots \\
    \lstick{\ket{\ell_{2}}} &  \qw & \ctrlo{-1} & \ctrlo{-1} & \ctrlo{-1}  & \ctrlo{-1}
    & \ctrl{-1} & \ctrl{-1}  & \ctrl{-1} & \ctrl{-1} & \qw & \cdots \\
    \lstick{\ket{\ell_{3}}} &  \qw & \ctrlo{-1} & \ctrlo{-1} & \ctrlo{-1} & \ctrlo{-1}
    & \ctrlo{-1} & \ctrlo{-1}  & \ctrlo{-1} & \ctrlo{-1} & \qw & \cdots \\
    \lstick{\ket{\ell_{4}}} &  \qw & \ctrlo{-1} & \ctrlo{-1} & \ctrlo{-1} & \ctrlo{-1}
    & \ctrlo{-1} & \ctrlo{-1}  & \ctrlo{-1} & \ctrlo{-1} & \qw & \cdots \\
    \lstick{\ket{\ell_{5}}} &  \qw & \ctrlo{-1} & \ctrlo{-1} & \ctrlo{-1} & \ctrlo{-1}
    & \ctrlo{-1} & \ctrlo{-1}  & \ctrlo{-1} & \ctrlo{-1} & \qw & \cdots \\
    \lstick{\ket{\vec{n}}} & {/} \qw & \ctrl{-1} & \ctrl{-1} & \ctrl{-1}  & \ctrl{-1} & \ctrl{-1} & \ctrl{-1} & \ctrl{-1} & \ctrl{-1} & \qw & \cdots
  }}
    \caption{The start of $O^{U(1)}_{A}$.  The control on $\ket{\vec{n}}$ is meant to indicate that each column must be controlled on, and its 32 nonzero elements inserted.}
    \label{fig:oau1}
\end{figure*}
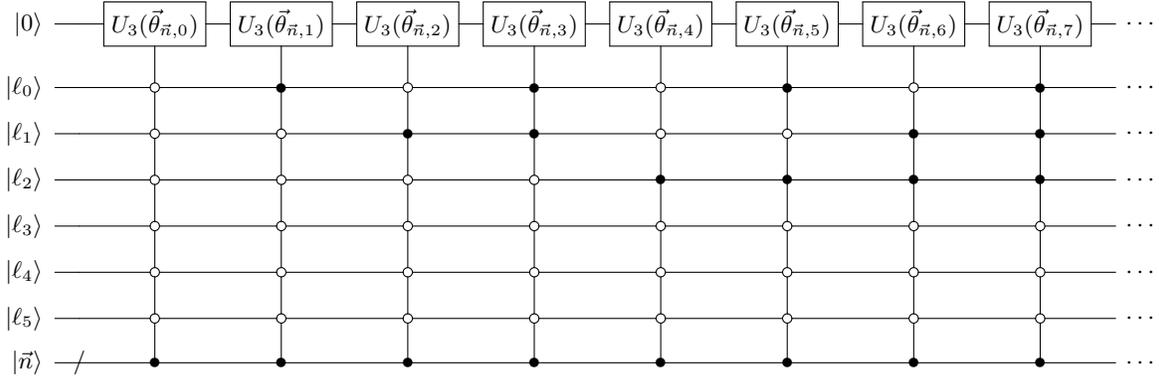

We now gauge the fermions with an abelian gauge field.
The action is,
\begin{align}
\nonumber
    S &= \sum_{n} \left\{ m \bar{\psi}(n) \psi(n) \right. \\ \nonumber
    &+ \left.  \frac{K}{2} \sum_{\mu=1}^{4} \eta_{\mu}(n) \left[ \bar{\psi}(n) U_{\mu}(n) \psi(n+\hat{\mu}) \right. \right. \\
    &- \left. \left. \bar{\psi}(n+\hat{\mu}) U^{*}_{\mu}(n) \psi(n) \right]  \right\}.
\end{align}
The fermion matrix is given by,
\begin{align}
  \label{eq:abel-stag-matrix}
  \nonumber
  M_{m n} &= \frac{K}{2} \sum_{\mu} \eta_{\mu}(m) (U_{\mu}(m) \delta_{n, m+\hat{\mu}} - U_{\mu}^{*}(m-\hat{\mu}) \delta_{n, m-\hat{\mu}}) \\
  &+ m_{0} \delta_{n, m}.
\end{align}
The resulting $W$ matrix is
\begin{align}
  \label{eq:abel-mdm-first}
  \nonumber
  W_{a c}  &= \frac{K^{2}}{4} \sum_{\mu, \nu} \eta_{\mu}(a) \eta_{\nu}(c) \left[
             U^{*}_{\mu}(a-\hat{\mu}) U_{\nu}(c-\hat{\nu}) \delta_{a-\hat{\mu}+\hat{\nu},c} \right. \\ \nonumber
  &-       U^{*}_{\mu}(a-\hat{\mu}) U^{*}_{\nu}(c) \delta_{a-\hat{\mu}-\hat{\nu},c} \\ \nonumber
  &-       U_{\mu}(a) U_{\nu}(c-\hat{\nu}) \delta_{a+\hat{\mu}+\hat{\nu},c} \\ \nonumber
           &+   \left.    U_{\mu}(a) U^{*}_{\nu}(c) \delta_{a+\hat{\mu}-\hat{\nu},c} \right] \\
  &+ m_{0}^{2} \delta_{a,c}.
\end{align}
and if we reorganize this equation to identify the unique NtNNs we find,
\begin{align}
  \label{eq:abel-mdm}
  W_{a c}  &= \frac{K^{2}}{4} \sum_{\mu < \nu}  [(\eta_{\mu}(a) \eta_{\nu}(c) U^{*}_{\mu}(a-\hat{\mu}) U_{\nu}(c-\hat{\nu}) \\ \nonumber
  &+ \eta_{\nu}(a) \eta_{\mu}(c) U_{\nu}(a) U^{*}_{\mu}(c)) \delta_{a-\hat{\mu}+\hat{\nu},c}  \\ \nonumber
           &-       (\eta_{\mu}(a) \eta_{\nu}(c) U^{*}_{\mu}(a-\hat{\mu}) U^{*}_{\nu}(c) \\ \nonumber
  &+ \eta_{\nu}(a) \eta_{\mu}(c) U^{*}_{\nu}(a-\hat{\nu}) U^{*}_{\mu}(c)) \delta_{a-\hat{\mu}-\hat{\nu},c} \\ \nonumber
           &-       (\eta_{\mu}(a) \eta_{\nu}(c) U_{\mu}(a) U_{\nu}(c-\hat{\nu}) \\ \nonumber
  &+ \eta_{\nu}(a) \eta_{\mu}(c) U_{\nu}(a) U_{\mu}(c-\hat{\mu})) \delta_{a+\hat{\mu}+\hat{\nu},c} \\ \nonumber
           &+ (\eta_{\nu}(a) \eta_{\mu}(c) U^{*}_{\nu}(a-\hat{\nu}) U_{\mu}(c-\hat{\mu})\\ \nonumber
  &+ \eta_{\mu}(a) \eta_{\nu}(c) U_{\mu}(a) U^{*}_{\nu}(c)) \delta_{a+\hat{\mu}-\hat{\nu},c} ] \\ \nonumber
           &+ (m_{0}^{2} + 2K^{2}) \delta_{a,c} \\ \nonumber
           &- \frac{K^{2}}{4} \sum_{\mu}(U^{*}_{\mu}(a-\hat{\mu}) U^{*}_{\mu}(a-2\hat{\mu}) \delta_{a-2\hat{\mu},c} \\ \nonumber
  &+ U_{\mu}(a) U_{\mu}(a+\hat{\mu}) \delta_{a+2\hat{\mu},c}).
\end{align}  
This matrix shows that each nonzero element on the off-diagonal consists of a sum over all paths from $c$ to $a$.
Since the nonzero elements of $W$ are between NtNNs, we must construct a $O^{NtNN}_{c}$ to take all these NtNNs into account.

There are 32 NtNNs and one diagonal element, meaning
we need to encode 33 nonzero elements in each column.  For this we require six qubits.  We use $\ell=0$-$31$ to encode the NtNNs, and $32\leq \ell < 64$ to encode the diagonal element.
We encode the two hops, starting with positive jumps, as
\begin{align}
  \label{eq:c-mdm-free}
  c(\vec{n},\ell) = 
  \begin{cases}
    \ell=0: & \vec{n}+2\hat{x} \\
    \ell=1: & \vec{n}+\hat{x}+\hat{y} \\
    \ell=2: & \vec{n}+\hat{x}+\hat{z} \\
    \ell=3: & \vec{n}+\hat{x}+\hat{t} \\
    \ell=4: & \vec{n}+2\hat{y} \\
    \ell=5: & \vec{n}+\hat{y}+\hat{z} \\
    \ell=6: & \vec{n}+\hat{y}+\hat{t} \\
    \ell=7: & \vec{n}+2\hat{z} \\
    \vdots & \vdots
  \end{cases}
\end{align}
Using the $\mathfrak{add}$ and $\mathfrak{sub}$ operators we can construct the $O^{NtNN}_{c}$ operator that connects NtNN pairs corresponding to Eq.~\eqref{eq:c-mdm-free}.  Looking at Eq.~\eqref{eq:abel-mdm},  the first term with $\sum_{\mu < \nu}$ can be done by combinations of $\mathfrak{add}$ and $\mathfrak{sub}$ matching those Kronecker deltas in the first term. This accounts for 24 NtNNs, the final eight come from the last term.  For this we use combinations of two $\mathfrak{add}$ or $\mathfrak{sub}$ operators for each direction.  Figure~\ref{fig:ocntnn} shows the circuit for $O^{NtNN}_{c}$.

Since each column of the matrix has 32 unique elements, in addition to the diagonal, we need to to specify which column we are populating with which matrix elements. This can be done by controlling on the $\ket{\vec{n}}$ register.  By controlling on $\ket{\vec{n}}$, along with the six qubits used for the 33 nonzero entries in each column, we can uniquely specify each nonzero matrix element.  We use the same encoding as $O^{NtNN}_{c}$ for the six qubits.  We define the action of $O_{A}^{U(1)}$ as,
\begin{align}
  O^{U(1)}_{A} \ket{0} \ket{\ell} \ket{\vec{n}} &= ( A_{c(\vec{n},\ell) \vec{n}} \ket{0} \\ \nonumber
 &+ \sqrt{1 - |A_{c(\vec{n},\ell) \vec{n}}|^{2}} \ket{1} ) \ket{\ell} \ket{\vec{n}}.
\end{align}
Each term in Eq.~\eqref{eq:abel-mdm} can be embedded in an SU(2) matrix, $U_{3}(\vec{\theta})$, as the top-left element.  For example, using the first term $\alpha \equiv  K^{2}(\eta_{\mu}(a) \eta_{\nu}(c) U^{*}_{\mu}(a-\hat{\mu}) U_{\nu}(c-\hat{\nu}) + \eta_{\nu}(a) \eta_{\mu}(c) U_{\nu}(a) U^{*}_{\mu}(c))/4$,
\begin{align}
  \label{eq:u3}
  U_{3} = 
  \begin{pmatrix}
    \alpha &-\sqrt{1 - |\alpha|^{2}} \\
    \sqrt{1 - |\alpha|^{2}} & \alpha^{*}
  \end{pmatrix}.
\end{align}
This SU(2) matrix can be parameterized by the three angles, $\vec{\theta}$, and directly decomposed in terms of elementary gates.
Just as before, $W$ must be normalized such that Eq.~\eqref{eq:u3} is unitary.  Moreover, $\alpha$ should be multiplied by the dimension of the Hilbert space of $\ket{\ell}$, to cancel the same factor that appears from the application of $D_{s}$ twice,
\begin{align}
  \label{eq:u3-final}
  U_{3} = 
  \begin{pmatrix}
    64 \alpha / s &-\sqrt{1 - |64 \alpha / s|^{2}} \\
    \sqrt{1 - |64 \alpha / s|^{2}} & 64 \alpha^{*} / s
  \end{pmatrix}.
\end{align}
The diagonal element---being real and the same in every column---can be encoded using a $y$-rotation just as in the free scalar field case. 
Figure~\ref{fig:oau1} shows the beginning of the quantum circuit for $O_{A}^{U(1)}$.

\subsection{SU(3) and staggered fermions}
\label{sec:su3-ferm}
We now gauge the fermions with the an SU(3) color gauge field.  The resulting fermion matrix has the same form as in the abelian case, with the minor addition of color indices.  The form of $W$ is identical to Eq.~\eqref{eq:abel-mdm-first} with the color indices expressed:
\begin{align}
  \label{eq:su3-mdm}
  \nonumber
  W^{\alpha \beta}_{a c}  &= \frac{K^{2}}{4} \sum_{\mu, \nu} \sum_{\gamma} \eta_{\mu}(a) \eta_{\nu}(c) \times \\ \nonumber
  &\left[
             {U^{\gamma \alpha}}^{*}_{\mu}(a-\hat{\mu}) U^{\gamma \beta}_{\nu}(c-\hat{\nu}) \delta_{a-\hat{\mu}+\hat{\nu},c} \right. \\ \nonumber
  &-       {U^{\gamma \alpha}}^{*}_{\mu}(a-\hat{\mu}) {U^{\beta \gamma}}^{*}_{\nu}(c) \delta_{a-\hat{\mu}-\hat{\nu},c} \\ \nonumber
  &-       U^{\alpha \gamma}_{\mu}(a) U^{\gamma \beta}_{\nu}(c-\hat{\nu}) \delta_{a+\hat{\mu}+\hat{\nu},c} \\ \nonumber
           &+   \left.    U^{\alpha \gamma}_{\mu}(a) {U^{\beta \gamma}}^{*}_{\nu}(c) \delta_{a+\hat{\mu}-\hat{\nu},c} \right] \\
  &+ m_{0}^{2} \delta_{a,c}.
\end{align}
The nonzero matrix elements are between NtNNs and so we can use $O^{NtNN}_{c}$. For $O^{SU(3)}_{A}$ we must encode a complex $3 \times 3$ matrix---as opposed to a sum of phases as in the abelian case---and so we need an additional four qubits to index the  matrix elements which we denote $\ell_{\alpha \beta} = 0,1,\ldots , 15$:
\begin{align}
  \nonumber
  O^{SU(3)}_{A} &\ket{0} \ket{\ell \ell_{\alpha \beta}} \ket{\vec{n}} = ( A^{\alpha \beta}_{c(\vec{n},\ell) \vec{n}} \ket{0} \\
 &+ \sqrt{1 - |A^{\alpha \beta}_{c(\vec{n},\ell) \vec{n}}|^{2}} \ket{1} ) \ket{\ell \ell_{\alpha \beta}} \ket{\vec{n}}.
\end{align}
$O_{A}^{SU(3)}$ can be constructed in a similar fashion as $O_{A}^{U(1)}$ with additional controls for $\ell_{\alpha \beta}$.
The above block-encoding enters each nonzero matrix element in $W$ into the quantum state which scales like $O(V)$, however each multi-control operates on $\log(V)$ qubits, and hence the complete scaling is $O(V\log(V))$.

\section{Matrix trace}
\label{sec:matrix-trace}
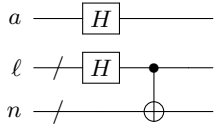
\begin{figure}
    \centering
    \mbox{
    \Qcircuit @C=1em @R=0.75em {
    \lstick{a} & \qw & \gate{H} & \qw & \qw & \qw & \\
    \lstick{\ell} & {/} \qw & \gate{H} & \ctrl{1} & \qw & \qw \\
    \lstick{n} & {/} \qw & \qw & \targ & \qw & \qw
    }}
    \caption{The state preparation circuit, $\prep$.  $H$ is the Hadamard gate.}
    \label{fig:state-prep}
\end{figure}

\begin{figure}
    \centering
    \mbox{
    \Qcircuit @C=1em @R=1em {
    \lstick{a} & \qw & \qw & \qw \\
    \lstick{\ell} & {/} \qw & \qw & \qw \\
    \lstick{n} & {/} \qw & \gate{\qop} & \qw
    }}
    \caption{The $Q$ matrix.}
    \label{fig:Q-matrix-circuit}
\end{figure}
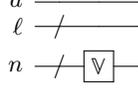

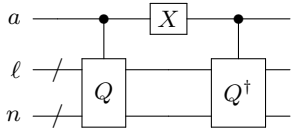
\begin{figure}
    \centering
    \mbox{
    \Qcircuit @C=1em @R=1em {
    \lstick{a} & \qw & \ctrl{1} & \gate{X} & \ctrl{1} & \qw \\
    \lstick{\ell} & {/} \qw & \multigate{1}{Q} & \qw & \multigate{1}{Q^{\dagger}} & \qw \\
    \lstick{n} & {/} \qw & \ghost{Q} & \qw & \ghost{Q^{\dagger}} & \qw 
    }}
    \caption{The $\mathbb{U}_{\varphi}$ matrix.  $X$ is the Pauli-$x$ matrix.}
    \label{fig:uphi}
\end{figure}

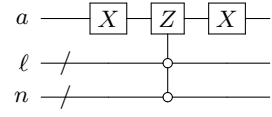
\begin{figure}
    \centering
    \mbox{
    \Qcircuit @C=1em @R=1em {
    \lstick{a} & \qw & \gate{X} & \gate{Z} & \gate{X} & \qw \\
    \lstick{\ell} & {/} \qw & \qw & \ctrlo{-1} & \qw & \qw\\
    \lstick{n} & {/} \qw & \qw & \ctrlo{-1} & \qw & \qw 
    }}
    \caption{The $-S_{0}$ circuit.  $Z$ and $X$ are the Pauli-$z$ and $x$ matrices, respectively.}
    \label{fig:allzeros}
\end{figure}

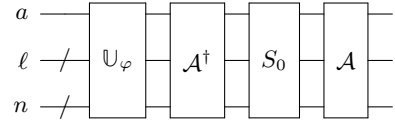
\begin{figure}
    \centering
    \mbox{
    \Qcircuit @C=1em @R=1em {
    \lstick{a} & \qw & \multigate{2}{\mathbb{U}_{\varphi}} & \multigate{2}{\prep^{\dagger}} & \multigate{2}{S_{0}} & \multigate{2}{\prep} & \qw \\
    \lstick{\ell} & {/} \qw & \ghost{\mathbb{U}_{\varphi}} & \ghost{\prep^{\dagger}} & \ghost{S_{0}} & \ghost{\prep} & \qw \\
    \lstick{n} & {/} \qw & \ghost{\mathbb{U}_{\varphi}} & \ghost{\prep^{\dagger}} & \ghost{S_{0}} & \ghost{\prep} & \qw
    }}
    \caption{The $\grov$ circuit.}
    \label{fig:grover}
\end{figure}

\begin{figure}
    \centering
    \includegraphics[width=8.6cm]{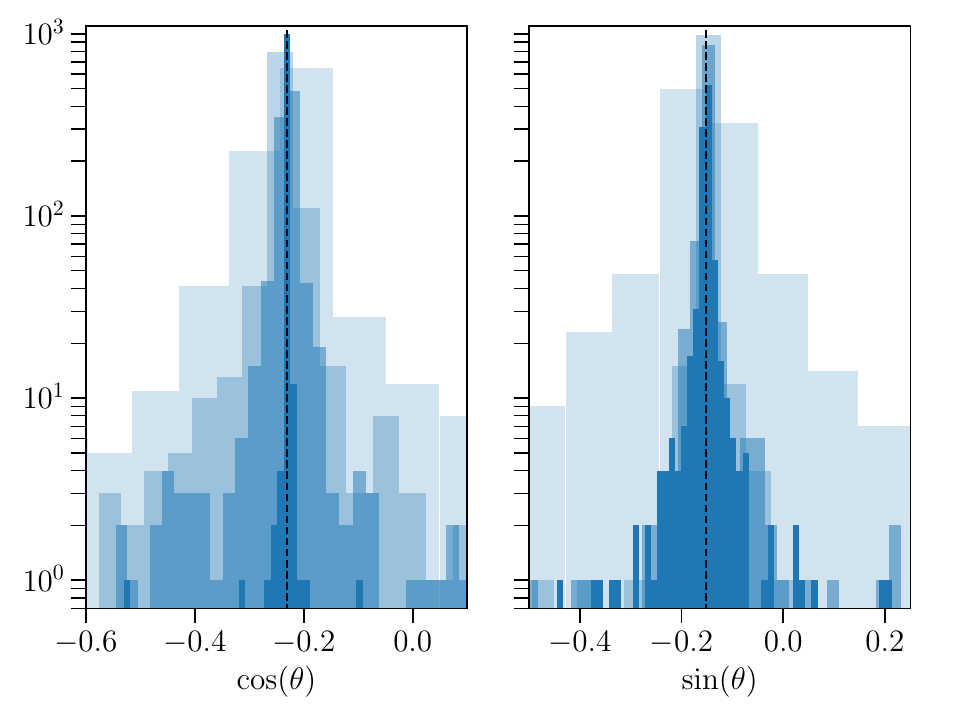}
    \caption{The shot counts of the real and imaginary parts of the trace of an example $4 \times 4$ unitary matrix for four different QPE qubit register sizes.  As the register size increases, the histogram bin widths decrease exponentially.  The darker histograms indicate a larger QPE register.  The black dashed lines indicate the exact values.}
    \label{fig:hist-converge}
\end{figure}

The next major ingredient to compute fermion determinants on a quantum computer is the matrix trace~\cite{PhysRevA.103.032422,aharonov2006polynomial,PhysRevLett.81.5672,cade2017quantum,shen2024efficient,Nghiem_2023}.  This can be done
by recasting the trace as an expectation value and using QME.  A similar classical approach can be carried out stochastically~\cite{Hutchinson,skorski2020modern}, and this approach has been pursued in the lattice QCD community for decades~\cite{PhysRevD.42.3520,DONG1994130}.  Quantum mean estimation uses quantum phase estimation (QPE).  To use QPE we need a state preparation method, and a unitary whose eigenphase corresponds to the average of interest.

We recast the trace of a $V \times V \equiv 2^{n} \times 2^{n}$ arbitrary unitary matrix, $U$, as an average through,
\begin{align}
\label{eq:trace-derivation}
\nonumber
\Gamma \equiv \frac{1}{V}\tr\left[U\right] &= \frac{1}{V}\sum_{i=0}^{V-1} U_{ii} \\ \nonumber
 &= \frac{1}{V}\sum_{i=0}^{V-1} \langle i\,|\, \qop\,|\,i\rangle\\ \nonumber
 &= \frac{1}{V}\sum_{i=0}^{V-1}\sum_{j=0}^{V-1} \delta_{ij}~\langle j\,|\,\qop\,|\,i\rangle\\ \nonumber
 &= \frac{1}{V}\sum_{i=0}^{V-1}\sum_{j=0}^{V-1} \langle j\,|\,i\rangle~~\langle j\,|\,\qop\,|\,i\rangle\\
 &= \langle \psi_0\,|\, Q \,|\,\psi_0\rangle\ \equiv \langle \psi_0\,|\,\mathbb{1} \otimes \qop\,|\,\psi_0\rangle\
 \end{align}
 where
 \begin{align}
 |\psi_0\rangle \equiv \frac{1}{\sqrt{V}}\sum_{i=0}^{V-1} |i\rangle \otimes |i\rangle,
\end{align}
and $\qop$ denotes the operator for $U$.  The form of Eq.~\eqref{eq:trace-derivation} is precisely the form of an average.  Moreover, since $Q$ is unitary, and $\ket{\psi_{0}}$ is a normalized state, QME can be used on a quantum computer to calculate it.

To use QME we first identify the various qubit registers.  We label them using the variable denoting their size.  First, we require a qubit register that handles the state space of $\qop$.  This is $n = \ceil{\log_{2}(V)}$ qubits.  Second, we need another register of equal size, $\ell = n$.  For QPE, we must set aside $m$ qubits which determine the precision and success probability, and finally QME as in Ref.~\cite{shyamsundar2021nonboolean} requires one additional qubit, which we shall label $a$.\footnote{If only the absolute value of the trace is desired, the $a$ qubit is unnecessary.}

State preparation for $|\psi_0\rangle$ uses $O(n)$ basic gates, and can be seen in Fig.~\ref{fig:state-prep}, along with the preparation of $a$.  We denoted the state preparation circuit as $\prep$.  The matrix $Q$ is given by $\mathbb{1} \otimes \qop$ in the $\ell$-$n$ state space.  The quantum circuit for $Q$ can be seen in Fig.~\ref{fig:Q-matrix-circuit}.  With this we can define the Grover iterate, $\grov$, used in QME.  Generically, $\grov =  \prep S_{0} \prep^{\dagger} \mathbb{U}_{\varphi}$, where $\mathbb{U}_{\varphi}$ is related to $Q$ through Fig.~\ref{fig:uphi}, and $S_{0}$ is the reflection about the all-zeros state, as shown in Fig.~\ref{fig:allzeros}.  The quantum circuit for $\grov$ is shown in Fig.~\ref{fig:grover}.  With the allotted qubits in register $m$, one uses $\prep$ and $\grov$ in QPE to extract a phase.  The cosine of this phase is precisely the real part of the average from Eq.~\eqref{eq:trace-derivation}.

To estimate the number of queries, we note that each call uses $O(n + n_{\qop})$ basic gates, where $n_{\qop}$ is the number of basic gates in $\qop$.  The
error in the estimate of the real part of the trace comes directly from QPE and is
\begin{align}
  \epsilon \sim O\left(\frac{1}{N_\text{calls}}\right),
\end{align}
which is the same as the imaginary part.  The success probability can be constrained to lie within $1-\delta$ with $N_{\text{calls}} \sim O(\log(1/\delta) / \epsilon)$~\cite{mande2023tight}.

To verify the algorithm, we compute the trace of an example $4 \times 4$ unitary matrix.  We do this for four different QPE qubit register sizes of $m=6,7,8,9$.  In this case $n = \ell = 2$.  The results of the shots from QPE for $\Re[\Gamma] \equiv \cos(\theta)$ and $\Im[\Gamma] \equiv \sin(\theta)$, corresponding to the real and imaginary parts of the trace, can be seen in Fig.~\ref{fig:hist-converge}, along with the exact values.  We find the algorithm converges as expected to the exact values.

\subsection{Trace of a block-encoding}
\label{sec:block-trace}
In actuality, we are interested in the trace of a block-encoded matrix.  A simple modification allows for this within the same framework of QME.
Consider the trace of the top-left  $V \times V$ block, $A$, in a block-encoding, $\qop$, where $\ell+1$ qubits are used to encode the block,
\begin{align}
    A = (\bra{0^{\ell+1}} \otimes \mathbb{1}) \qop (\ket{0^{\ell+1}} \otimes \mathbb{1}).
\end{align}
Then,
\begin{align}
\label{eq:block-trace}
\nonumber
 frac{1}{V}\tr\left[A\right] &= \frac{1}{V}\sum_{i=0}^{V-1} A_{ii} \\ \nonumber
 &= \frac{1}{V}\sum_{i=0}^{V-1} \bra{0^{\ell+1}} \bra{i} \qop \ket{0^{\ell+1}} \ket{i}\\ \nonumber
 &= \frac{1}{V} \sum_{j=0}^{V-1} \sum_{i=0}^{V-1} \delta_{ij} \bra{0^{\ell+1}} \bra{j} \qop \ket{0^{\ell+1}} \ket{i}\\ \nonumber
  &= \frac{1}{V} \sum_{j=0}^{V-1} \sum_{i=0}^{V-1} \braket{j | i} \bra{0^{\ell+1}} \bra{j} \qop \ket{0^{\ell+1}} \ket{i}\\
  &= \bra{\psi_{0}} Q  \ket{\psi_{0}} \equiv \bra{\psi_{0}} \mathbb{1} \otimes \qop \ket{\psi_{0}}
 \end{align}
 with
 \begin{align}
     \ket{\psi_{0}} = \frac{1}{\sqrt{V}} \sum_{i} \ket{i} \ket{0^{\ell+1}} \ket{i}
 \end{align}
 and $\mathbb{1}$ is the identity on a Hilbert space with the same dimension as the block.  With this minor modification to the initial state
 used in QME we can compute the trace of a block of a matrix.

\section{Determinant of a matrix}
\label{sec:matrix-det}
\begin{figure*}
  \centering
  \includegraphics[width=17.2cm]{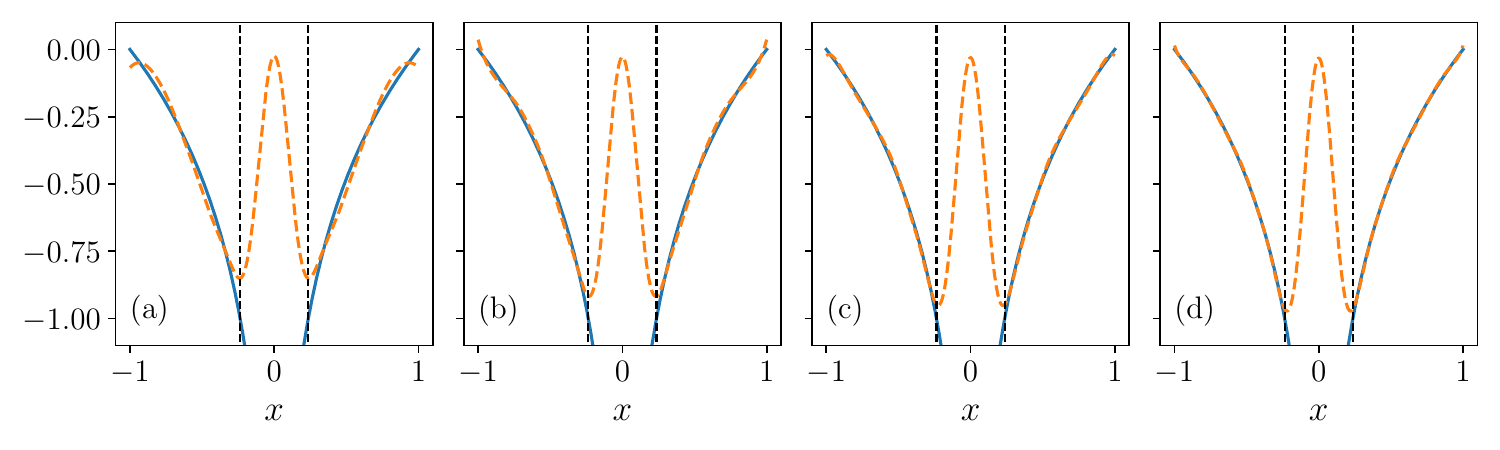}
  \caption{A comparison of polynomials with increasingly larger degree. The vertical black dashed lines indicate an example of the location of the smallest eigenvalue of $W$, which is where $\log(|x|) / \log(1/\lambda_{\text{min}}) = -1$.  The blue line is the exact function $\log(|x|) / \log(1/\lambda_{\text{min}})$, and the orange line is the polynomial approximation of degree (a) $d=64$, (b) $d=66$, (c) $d=68$, and (d) $d=70$. Here $d_{r} = 60$ was used in all four cases.}
  \label{fig:d-compare}
\end{figure*} 

We are now ready to combine the results from the previous sections and compute the determinant of $W$ as defined in Sec.~\ref{sec:su3-ferm}.
To compute the determinant we use the relation $\Tr[\log(W)] = \log\det(W)$.  We do this because it is easy on a quantum computer to compute the $\log$ of a matrix using the QET, and the trace can be computed using the method from Sec.~\ref{sec:block-trace}.

The QET is possible because of the following theorem of quantum signal processing~\cite{Gily_n_2019}:
\begin{theorem}
  Quantum signal processing in SU(2): For any polynomials $P$, $Q \in \mathbb{C}$, and any positive integer $d$, such that (1) $\deg(P) \leq d$, $\deg(Q) \leq (d-1)$, (2) $P$ has parity $d \mod 2$, and $Q$ has parity $(d-1) \mod 2$, (3) $|P(x)|^{2} + (1-x^{2})|Q(x)|^{2} = 1$, $\forall x \in [-1,1]$, there exists a set of phase factors $\Phi = \{\phi_{0}, \ldots, \phi_{d} \} \in [-\pi, , \pi)^{d+1}$ such that
  \begin{align}
    \label{eq:qsp}
    U_{\Phi} &= e^{i \phi_{0} Z} \prod_{n=1}^{d} W(x) e^{i \phi_{n} Z} \\
             &=
               \begin{pmatrix}
                 P(x) & i Q(x) \sqrt{1-x^{2}} \\
                 i Q(x) \sqrt{1-x^{2}} & P^{*}(x)
               \end{pmatrix},
  \end{align}
  where
  \begin{align}
    \label{eq:W-def}
    W(x) = e^{i \arccos(x) X} =
    \begin{pmatrix}
      x & i \sqrt{1-x^{2}} \\
      i \sqrt{1-x^{2}} & x
    \end{pmatrix}.
  \end{align}
\end{theorem}
If one can find the appropriate phases for a polynomial approximation of $\log(x)$, then the QET can use those phases to apply that same polynomial approximation to a block-encoded matrix.

Under the polynomial transform, the resulting polynomial must be between $-1$ and $1$ since it ultimately is used in a unitary block-encoding. So if $W$ is rescaled such that the largest eigenvalue $\lambda_{\text{max}} \leq 1$, and $\lambda_{\text{min}}$ is the smallest eigenvalue of $W$, then $0 \geq \log(W) / \log(1/\lambda_{\text{min}}) \geq -1$, and we should find a polynomial approximation, $\tilde{f}$, of $f(x) \equiv \log(x) / \log(1/\lambda_{\text{min}})$.
This additional normalization of $W$ is compatible with the normalization necessary for the block-encoding.  The largest eigenvalue of $W$ is bounded by $m_{0}^{2} \leq \lambda_{\text{max}} \leq m_{0}^{2} + 16K^{2}$ (see appendix~\ref{subsec:app:eig-bound}).
To give our function definite parity we actually define $f(x) \equiv \log(|x|) / \log(1/\lambda_{\text{min}})$, which changes nothing since we are interested in cases where $x > 0$.

To approximate $f$, we use a Chebyshev series,
\begin{align}
  \label{eq:f-cheby}
  \tilde{f}(x) = \sum_{n=0}^{d_{f}} a_{n} T_{n}(x) \approx f(x),
\end{align}
over the interval $\lambda_{\text{min}} \leq |x| \leq 1$, keeping only even terms. This polynomial approximation can however be less than negative one, in particular when $|x| < \lambda_{\text{min}}$.  To fix this we use the same procedure from Ref.~\cite{Gily_n_2019}, and construct a polynomial approximation to the rectangular function,
\begin{align}
  \label{eq:rect-func}
  r(x, s) =
  \begin{cases}
    1 & |x| > s \\
    0 & |x| < s \\
    \frac{1}{2} & |x| = s
  \end{cases}.
\end{align}
This is done by first taking the error function as a proxy for the sign function, and then adding two shifted error functions to form an approximate rectangular function.  Polynomial approximations for such functions can be found in Ref.~\cite{Gily_n_2019}.  We find performing a Chebyshev fit to the error-function approximation of the rectangular function works better in practice, keeping only even terms.
We denote this polynomial approximation of the rectangular function as $\tilde{r}(x,s)$.  $\tilde{r}$ has degree $d_{r}$.
Since the product of two polynomials is another polynomial, we can construct an improved polynomial approximation to $f$, $F$, using the product $F(x) \equiv \tilde{r}(x,\lambda_{\text{min}}) \times \tilde{f}(x)$ with degree $d = d_{f} + d_{r}$.  We then use $F$ for the polynomial $P$ found in Eq.~\eqref{eq:qsp}.

The phases can be found using the method of least squares~\cite{PhysRevA.103.042419}.    Using $F$ from above, along with the method from Ref.~\cite{PhysRevA.103.042419}, we are able to find phases such that the $P(x)$ from Eq.~\eqref{eq:qsp} reproduces $F$ to machine precision. Figure~\ref{fig:d-compare} shows a comparison between the exact function $f(x)$ and the polynomial approximation found using the phases obtained from the least squares minimization.  We find good agreement by setting the degree of the rectangular polynomial much higher than that of the logarithm polynomial.  In addition, we find it helps to shift the entire polynomial away from $y=-1$, so that the polynomial ``sits'' in the middle of 1 and -1.  A shift of $1/2$ works well.  This known shift can always be subtracted at the end of the calculation.

With a block-encoding of the matrix (from Sec.~\ref{sec:block-encoding}) to which we wish to apply the polynomial transform using the phases, we can use the method of alternating phase modulation~\cite{Gily_n_2019,low2024quantum}
and perform the QET to approximate $F$ in the uppermost-left block of the unitary block-encoding.
We then compute the trace of this matrix using the method from Sec.~\ref{sec:block-trace}, yielding $\approx \log\det(W) / \log(1/\lambda_{\text{min}})$. We also note that this sequence of operations is gauge invariant. Since the QET applies a polynomial transform on the eigenvalues of the block-encoded matrix, and a gauge transform leaves the eigenvalues of the fermion matrix invariant, the final result is gauge invariant.  We checked this explicitly with an example gauge field to verify and find it holds true.

\section{Conclusions}
\label{sec:conclusion}
We have presented an algorithm to compute the logarithm of the determinant of the staggered fermion matrix for a classical gauge field configuration.  The algorithm uses the relation $\Tr[\log] = \log\det$ and computes the logarithm of the fermion matrix using the QET.  The trace is them computed using QME.  We find an improved scaling with this quantum algorithm relative to the classical computing algorithms which are either exact
based on sparse LU decomposition \cite{wang2023accelerating} to compute the determinant,
or stochastic trace estimation \cite{Hutchinson}; with the present algorithm scaling like $O(V \log(V))$ while the traditional classical algorithm scales like $O(V^{3})$.

The $O(V \log(V))$ scaling stems from the quantum circuit cost associated with the block-encoding of the classical data.  This quantum circuit is then run through the QET $O(d)$ times. This is in contrast to a classical version of this algorithm where one would construct a matrix polynomial of $W$ that approximates the logarithm.
Classically, the cost of computing a polynomial of $W$ is $O(d\,V^3)$, demonstrating a polynomial speed-up when using the quantum algorithm. The $O(V^3)$ scaling arises from matrix multiplication operations, even in the case of an initial sparse matrix.
A detailed analysis of the block-encoding complexity can be found in appendix~\ref{app:complexity}.

An important ingredient of the algorithm is the use of the fact that $2 \log\det(M) = \log\det(W)$, which allows the computation of $\Tr\log(W)$ in place of $M$.  The $W$ matrix actually contains an additional property not taken advantage of here; that is that $W$ only couples NtNN lattice sites, meaning that only half the matrix need be block-encoded, since the even and odd sub-lattices are completely independent of each other.  We save this optimization for future work.  Relevantly, future work also includes devising a similar algorithm for Wilson and overlap/domain wall fermions.

With the success of the QET to compute polynomial transforms of matrices, it would be interesting to extend the ideas presented here to other classical algorithms used in lattice gauge theory calculations, \emph{e.g.} matrix inversion and the trace operation in the calculation of the chiral condensate.  Another possible benefit would be the calculation of the sign of the Hermitian Wilson-Dirac operator in QCD which is currently an expensive operation, but which could perhaps benefit from the QET's speed-up.

\begin{acknowledgments}
This manuscript has been authored by employees of Fermi Research Alliance, LLC under Contract No. DE-AC02-07CH11359 with the U.S. Department of Energy, Office of Science, Office of High Energy Physics.  This work is supported by the Department of Energy through the Fermilab Theory QuantiSED program in the area of ``Intersections of QIS and Theoretical Particle Physics''. PS is supported by the U.S. Department of Energy, Office of Science, Office of High Energy Physics QuantISED program under the grants ``HEP Machine Learning and Optimization Go Quantum'', Award Number 0000240323, and ``DOE QuantiSED Consortium QCCFP-QMLQCF'', Award Number DE-SC0019219.
\end{acknowledgments}

\appendix

\section{Fermion matrix eigenvalue relations}
\label{sec:app:ferm-eig-relations}
Write the fermion matrix as
\begin{align}
  M = \slashed{D} + m_{0}.
\end{align}
The spectrum, $\sigma$, of this operator is characterized by
\begin{align}
  \label{eq:m-spectrum}
  \sigma(M) = \{ \pm i\lambda + m_{0}: \pm i\lambda \in \sigma(\slashed{D}) \}.
\end{align}
We have that
\begin{align}
 W &= M^\dagger M\,,\\
 \Longrightarrow \det\left[W\right] &= \det\left[M^\dagger\right] \det\left[M\right]\\
 &= \left(\det\left[M\right]\right)^\ast \det\left[M\right] = \Big|\det\left[M\right]\Big|^2
\end{align}
Since the eigenvalues of $M$ occur in complex conjugate pairs, $\det[M]$ is real-valued and non-negative. So,
\begin{align}
 \det\left[M\right] = \sqrt{\det\left[W\right]}\Rightarrow \tr\left[\log M\right] = \frac{1}{2}\tr\left[\log W\right].
\end{align}

\section{Spectral bounds on staggered Dirac operator}
\label{subsec:app:eig-bound}
We write the staggered Dirac operator in the following form
\begin{align}
  \label{eq:app:M}
  M_{x x'} &= N \left( 2 m' \delta_{x, x'} \right. \\ \nonumber
             &+ \left.  \sum_{\mu = 1}^{4} \eta_{\mu}(x) \left[ U_{\mu}(x) \delta_{x+\hat{\mu},x'} - U^{\dagger}_{\mu}(x-\hat{\mu}) \delta_{x-\hat{\mu},x'} \right] \right),
\end{align}
where $N$ is an arbitrary normalization, related to Eq.~\eqref{eq:abel-stag-matrix} by $N \equiv K / 2$ and $m' = m_{0}/K$.
For this analysis $\eta_{\mu}^{*}(x-\hat{\mu}) = \eta_{\mu}(x)$.  In the beginning we will work with the normalization $N=1$ and mention how the spectral bound changes (trivially) with normalization at the end.

To proceed, we define a set of four unitary matrices
\begin{align}
  \label{eq:app:bigU}
  \mathcal{U}_{\mu} \equiv \eta_{\mu}(x) U_{\mu}(x) \delta_{x+\hat{\mu}, x'}.
\end{align}
and rewrite the staggered Dirac operator ($N=1$) in terms of  these matrices
\begin{align}
  M = 2m' + \sum_{\mu} (\mathcal{U}_{\mu} - \mathcal{U}^{\dagger}).
\end{align}
For a given $\mu$ we have the following eigenvalue equations
\begin{align}
  \mathcal{U}_{\mu} \ket{\mu, j} &= e^{i \theta_{\mu, j}} \ket{\mu, j} \\ \nonumber
  (\mathcal{U}_{\mu} - \mathcal{U}^{\dagger}_{\mu}) \ket{\mu, j} &= (e^{i \theta_{\mu, j}} - e^{-i \theta_{\mu, j}}) \ket{\mu, j} \\
  &= 2i \sin(\theta_{\mu, j}) \ket{\mu, j}
\end{align}
For normal eigenvectors, we see the bounds are
\begin{align}
  |\braket{\mu, j |\mathcal{U}_{\mu} | \mu, j}| = 1 \\
  |\braket{\mu, j | (\mathcal{U}_{\mu} - \mathcal{U}^{\dagger}_{\mu}) | \mu, j}| \leq 2 .
\end{align}
In general $[\mathcal{U}^{\dagger}_{\mu}, \mathcal{U}_{\nu}] \neq \delta_{\mu \nu}$, so it is not so simple to find a basis where $\sum_{\mu} (\mathcal{U}_{\mu} - \mathcal{U}^{\dagger}_{\mu})$ is diagonal.  However, the bounds are still additive due to an analogue of the Schwartz inequality.  So, we get the following spectral bound for the mass independent part of $M$
\begin{align}
 \left[ \textstyle{\sum_{\mu=1}^4} (\mathcal{U}_{\mu} - \mathcal{U}^{\dagger}_{\mu}) \right]
 \ket{\lambda} = \lambda \ket{\lambda}, \quad |\lambda| \leq 8.
\end{align}
Because $M$ is anti-Hermitian when $m'=0$, we also get the following spectral bound on $M$
\begin{align}
  M \ket{\lambda'} = \lambda' \ket{\lambda'}, \quad 2|m'| \leq \lambda' \leq \sqrt{4{m'}^{2} + 64}.
\end{align}
For the Hermitian positive-definite form $W = M^{\dagger} M$, the spectrum is real and positive with the spectral bound
\begin{align}
  W \ket{\lambda''} = \lambda'' \ket{\lambda''}, \quad 4 {m'}^{2} \leq \lambda'' \leq 4{m'}^{2} + 64.
\end{align}
We trivially extend this result to general normalizations of the staggered  Dirac operator
\begin{align}
  \nonumber
  &W \ket{\lambda''} = \lambda'' \ket{\lambda''}, \quad 4 N^{2} {m'}^{2} \leq \lambda'' \leq 4N^{2} ({m'}^{2} + 16) \\
  &= m_{0}^{2} \leq \lambda'' \leq  ({m_{0}}^{2} + 16 K^{2}).
\end{align}
Also, we note that the upper limit on the condition number $\kappa$ for $W$  is independent of normalization
\begin{align}
  \kappa \leq \frac{16 + {m'}^{2}}{{m'}^{2}} \approx \frac{16}{{m'}^{2}} = \frac{16 K^{2}}{m_{0}^{2}}
\end{align}
where we assume $m' \ll 1$.

\section{Computational complexity}
\label{app:complexity}
For completeness, here we list the computational complexity in terms of the spacetime dimension $D$, and volume $V$, for
the various parts of the above algorithms for block-encoding in table~\ref{tab:simple_table}.  The computational complexity
for the QET is well established, and can be found in the original references~\cite{PRXQuantum.2.040203,low2024quantum,Gily_n_2019}, and
for QME it is the same as quantum phase estimation~\cite{mande2023tight}. The overall computational complexity of our algorithm is the product of complexities of the block encoding, QET, and QME. Importantly, computational complexities of QET and QME, for a given target error bound, are independent of $V$.

\begin{table*}[h!]
\centering
\begin{tabular}{c c c c c c c}
  \toprule[1.2pt]
  & \multicolumn{2}{c}{Number of qubits} &\multicolumn{4}{c}{Number of gates (for parts of the block encoding circuit)}\\
  \cmidrule(lr){2-3}
  \cmidrule(lr){4-7}
  & $\ket{\vec{n}}$ register & $\ket{\ell}$ register  & $D_s$
  & $O_c$ & $O_A$ & overall \\
  \midrule[1.2pt]
  scalar Laplacian & $O(\log V)$ & $O(\log D)$ & $O(\log D)$ & $O(D\,\log D\,\log V)$ & $O(1)$ & $O(D\,\log D\,\log V)$ \\
  \cmidrule(lr){1-7}
  free fermions & $O(\log V)$ & $O(\log D)$ & $O(\log D)$ &  $O(D\,\log D\,\log V)$ & $O(1)$ & $O(D\,\log D\,\log V)$ \\
  \cmidrule(lr){1-7}
  U(1) & $O(\log V)$ & $O(\log D)$ & $O(\log D)$ & $O(D^{2}\,\log D\,\log V)$ & $O(D^{2}\,V\,\log(D\,V))$ & $O(D^{2}\,V\,\log(D\,V))$ \\
  \cmidrule(lr){1-7}
  SU(3) & $O(\log V)$ & $O(\log D)$ & $O(\log D)$ & $O(D^{2}\,\log D\,\log V)$ & $O(D^{2}\,V\,\log(D\,V))$ & $O(D^{2}\,V\,\log(D\,V))$ \\
  \bottomrule[1.2pt]
\end{tabular}
\caption{The qubits and computational complexity for the different block-encodings explained in the main text.}
\label{tab:simple_table}
\end{table*}

\bibliographystyle{unsrt}

\begin{thebibliography}{10}

\bibitem{montvay_munster_1994}
Istvan Montvay and Gernot M{\"{u}}nster.
\newblock {\em Quantum Fields on a Lattice}.
\newblock Cambridge Monographs on Mathematical Physics. Cambridge University Press, 1994.

\bibitem{gattringer}
Christof Gattringer and Christian~B. Lang.
\newblock {\em Quantum Chromodynamics on the Lattice}.
\newblock Lecture Notes in Physics. Springer Berlin, Heidelberg, 2010.

\bibitem{kogut:1979}
John~B. Kogut.
\newblock An introduction to lattice gauge theory and spin systems.
\newblock {\em Rev. Mod. Phys.}, 51:659--713, Oct 1979.

\bibitem{WEINGARTEN1981333}
D.H. Weingarten and D.N. Petcher.
\newblock Monte carlo integration for lattice gauge theories with fermions.
\newblock {\em Physics Letters B}, 99(4):333--338, 1981.

\bibitem{Gottlieb:1987mq}
Steven~A. Gottlieb, W.~Liu, D.~Toussaint, R.~L. Renken, and R.~L. Sugar.
\newblock {Hybrid Molecular Dynamics Algorithms for the Numerical Simulation of Quantum Chromodynamics}.
\newblock {\em Phys. Rev.}, D35:2531--2542, 1987.

\bibitem{Joo:2001bz}
B.~Joo, I.~Horvath, and K.~F. Liu.
\newblock {The Kentucky noisy Monte Carlo algorithm for Wilson dynamical fermions}.
\newblock {\em Phys. Rev. D}, 67:074505, 2003.

\bibitem{PhysRevD.105.L051506}
Szabolcs Bors\'anyi, Zolt\'an Fodor, Matteo Giordano, S\'andor~D. Katz, D\'aniel N\'ogr\'adi, Attila P\'asztor, and Chik~Him Wong.
\newblock Lattice simulations of the qcd chiral transition at real baryon density.
\newblock {\em Phys. Rev. D}, 105:L051506, Mar 2022.

\bibitem{Nagata_2022}
Keitaro Nagata.
\newblock Finite-density lattice qcd and sign problem: Current status and open problems.
\newblock {\em Progress in Particle and Nuclear Physics}, 127:103991, November 2022.

\bibitem{Park:2024vjp}
Sungwoo Park, Tanmoy Bhattacharya, Rajan Gupta, Huey-Wen Lin, Santanu Mondal, and Boram Yoon.
\newblock {Update on flavor diagonal nucleon charges from clover fermions}.
\newblock {\em PoS}, LATTICE2023:328, 2024.

\bibitem{Kuberski:2023qgx}
Simon Kuberski.
\newblock {Muon $g-2$: Lattice calculations of the hadronic vacuum polarization}.
\newblock {\em PoS}, LATTICE2023:125, 2024.

\bibitem{Ellis:2018dmb}
John Ellis, Natsumi Nagata, and Keith~A. Olive.
\newblock {Uncertainties in WIMP Dark Matter Scattering Revisited}.
\newblock {\em Eur. Phys. J. C}, 78(7):569, 2018.

\bibitem{Varnhorst:2020dba}
Lukas Varnhorst.
\newblock {\em {Aspects of quark mass dependence in lattice QCD}}.
\newblock PhD thesis, Wuppertal U., 2020.

\bibitem{LatticeStrongDynamics:2023bqp}
R.~C. Brower et~al.
\newblock {Light Scalar Meson and Decay Constant in SU(3) Gauge Theory with Eight Dynamical Flavors}.
\newblock 6 2023.

\bibitem{PRXQuantum.2.040203}
John~M. Martyn, Zane~M. Rossi, Andrew~K. Tan, and Isaac~L. Chuang.
\newblock Grand unification of quantum algorithms.
\newblock {\em PRX Quantum}, 2:040203, Dec 2021.

\bibitem{low2024quantum}
Guang~Hao Low and Yuan Su.
\newblock Quantum eigenvalue processing, 2024.

\bibitem{Gily_n_2019}
András Gilyén, Yuan Su, Guang~Hao Low, and Nathan Wiebe.
\newblock Quantum singular value transformation and beyond: exponential improvements for quantum matrix arithmetics.
\newblock In {\em Proceedings of the 51st Annual ACM SIGACT Symposium on Theory of Computing}, STOC ’19. ACM, June 2019.

\bibitem{kothari:2022}
Robin Kothari and Ryan O'Donnell.
\newblock {Mean estimation when you have the source code; or, quantum Monte Carlo methods}, 8 2022.

\bibitem{shyamsundar2021nonboolean}
Prasanth Shyamsundar.
\newblock Non-boolean quantum amplitude amplification and quantum mean estimation, 2021.

\bibitem{Ham21}
Yassine Hamoudi.
\newblock {\em Quantum Algorithms for the Monte Carlo Method}.
\newblock PhD thesis, Universit\'{e} de Paris, 2021.

\bibitem{montanaro:2015}
Ashley Montanaro.
\newblock Quantum speedup of monte carlo methods.
\newblock {\em Proceedings of the Royal Society A: Mathematical, Physical and Engineering Sciences}, 471(2181):20150301, 2015.

\bibitem{Low_2019}
Guang~Hao Low and Isaac~L. Chuang.
\newblock Hamiltonian simulation by qubitization.
\newblock {\em Quantum}, 3:163, July 2019.

\bibitem{PhysRevLett.118.010501}
Guang~Hao Low and Isaac~L. Chuang.
\newblock Optimal hamiltonian simulation by quantum signal processing.
\newblock {\em Phys. Rev. Lett.}, 118:010501, Jan 2017.

\bibitem{camps2023explicit}
Daan Camps, Lin Lin, Roel~Van Beeumen, and Chao Yang.
\newblock Explicit quantum circuits for block encodings of certain sparse matrices, 2023.

\bibitem{PhysRevA.103.032422}
Anirban~N. Chowdhury, Rolando~D. Somma, and Yi\ifmmode \breve{g}\else~\u{g}\fi{}it Suba\ifmmode \mbox{\c{s}}\else \c{s}\fi{}\ifmmode \imath \else~\i \fi{}.
\newblock Computing partition functions in the one-clean-qubit model.
\newblock {\em Phys. Rev. A}, 103:032422, Mar 2021.

\bibitem{aharonov2006polynomial}
Dorit Aharonov, Vaughan Jones, and Zeph Landau.
\newblock A polynomial quantum algorithm for approximating the jones polynomial, 2006.

\bibitem{PhysRevLett.81.5672}
E.~Knill and R.~Laflamme.
\newblock Power of one bit of quantum information.
\newblock {\em Phys. Rev. Lett.}, 81:5672--5675, Dec 1998.

\bibitem{cade2017quantum}
Chris Cade and Ashley Montanaro.
\newblock The quantum complexity of computing schatten $p$-norms, 2017.

\bibitem{shen2024efficient}
Yizhi Shen, Katherine Klymko, Eran Rabani, Daan Camps, Roel~Van Beeumen, and Michael Lindsey.
\newblock Efficient quantum trace estimation with reconfigurable real-time circuits, 2024.

\bibitem{Nghiem_2023}
Nhat~A. Nghiem and Tzu-Chieh Wei.
\newblock An improved method for quantum matrix multiplication.
\newblock {\em Quantum Information Processing}, 22(8), August 2023.

\bibitem{Hutchinson}
M.F. Hutchinson.
\newblock A stochastic estimator of the trace of the influence matrix for laplacian smoothing splines.
\newblock {\em Communications in Statistics - Simulation and Computation}, 18(3):1059--1076, 1989.

\bibitem{skorski2020modern}
Maciej Skorski.
\newblock A modern analysis of hutchinson's trace estimator, 2020.

\bibitem{PhysRevD.42.3520}
H.~R. Fiebig and R.~M. Woloshyn.
\newblock Monopoles and chiral-symmetry breaking in three-dimensional lattice qed.
\newblock {\em Phys. Rev. D}, 42:3520--3523, Nov 1990.

\bibitem{DONG1994130}
Shao-Jing Dong and Keh-Fei Liu.
\newblock Stochastic estimation with z2 noise.
\newblock {\em Physics Letters B}, 328(1):130--136, 1994.

\bibitem{mande2023tight}
Nikhil~S. Mande and Ronald de~Wolf.
\newblock Tight bounds for quantum phase estimation and related problems, 2023.

\bibitem{PhysRevA.103.042419}
Yulong Dong, Xiang Meng, K.~Birgitta Whaley, and Lin Lin.
\newblock Efficient phase-factor evaluation in quantum signal processing.
\newblock {\em Phys. Rev. A}, 103:042419, Apr 2021.

\bibitem{wang2023accelerating}
Tengcheng Wang, Wenhao Li, Haojie Pei, Yuying Sun, Zhou Jin, and Weifeng Liu.
\newblock Accelerating sparse lu factorization with density-aware adaptive matrix multiplication for circuit simulation.
\newblock In {\em 2023 60th ACM/IEEE Design Automation Conference (DAC)}, pages 1--6. IEEE, 2023.

\end{thebibliography}

\end{document}